\documentclass[preprint]{aastex631}

\usepackage{amsmath}
\usepackage{bm}
\usepackage{verbatim}

\shorttitle{Sgr High-velocity stars}
\shortauthors{Li et al.}

\begin{document}

\title{60 candidate high-velocity stars originating from the Sagittarius dwarf spheroidal galaxy in Gaia EDR3}

\correspondingauthor{Cuihua Du}
\email{ducuihua@ucas.ac.cn}

\author{Hefan Li}
\affiliation{School of Physical Sciences, University of Chinese Academy of Sciences, Beijing 100049, P. R. China}

\author{Cuihua Du}
\affiliation{College of Astronomy and Space Sciences, University of Chinese Academy of Sciences, Beijing 100049, P.R. China}

\author{Jun Ma}
\affiliation{Key Laboratory of Optical Astronomy, National Astronomical Observatories, Chinese Academy of Sciences, Beijing 100012, P.R.China}
\affiliation{College of Astronomy and Space Sciences, University of Chinese Academy of Sciences, Beijing 100049, P.R. China}

\author{Jianrong Shi}
\affiliation{Key Laboratory of Optical Astronomy, National Astronomical Observatories, Chinese Academy of Sciences, Beijing 100012, P.R.China}
\affiliation{College of Astronomy and Space Sciences, University of Chinese Academy of Sciences, Beijing 100049, P.R. China}

\author{Heidi Jo Newberg}
\affiliation{Department of Physics, Applied Physics and Astronomy, Rensselaer Polytechnic Institute, Troy, NY 12180, USA}

\author{Yunsong Piao}
\affiliation{School of Physical Sciences, University of Chinese Academy of Sciences, Beijing 100049, P. R. China}

\begin{abstract}
\par Using proper motions from Gaia Early Data Release 3 (Gaia EDR 3) and radial velocities from several surveys, we identify 60 candidate high-velocity stars with total velocity greater than 75\% escape velocity that probably origin from Sagittarius dwarf spheroidal galaxy (Sgr) by orbital analysis. Sgr's gravity has little effect on the results and the Large Magellanic Cloud's gravity has non-negligible effect on only a few stars. The closest approach of these stars to the Sgr occurs when the Sgr passed its pericenter ($\sim$ 38.2 Myr ago), which suggest they were tidally stripped from the Sgr. The positions of these stars in the HR diagram and the chemical properties of 19 of them with available [Fe/H] are similar with the Sgr stream member stars. This is consistent with the assumption of their accretion origin. Two of the 60 are hypervelocity stars, which may also be produced by Hills mechanism.
\end{abstract}


\section{introduction}

\par High-velocity stars move so fast that it implies the presence of extreme dynamical processes. One type of high-velocity star is hypervelocity star (HVS), which has a extreme velocity above the escape velocity of the Milky Way. \citet{hills88Hypervelocity} first theoretically predicted the formation of HVS by close encounter between a binary system and the massive black hole (MBH) in the Galactic Center (GC). \citet{yu03Ejection} analyze the ejection mechanism of single stars and a binary MBH and calculate the ejection rate. The first candidate HVS was discovered by \citet{brown05Discovery}: a B-type star with a total velocity of 709 km s$^{-1}$ in the Galactic halo. Recently, the Southern Stellar Stream Spectroscopic Survey ($S^5$) discovered the currently fastest HVS \citep{koposov20Discovery}: an A-type star with a total velocity of 1755 km s$^{-1}$ located at a heliocentric distance of $\sim 9$ kpc, and its backward orbit was found to point unambiguously to the GC, providing a direct evidence to the Hills mechanism.  
 
\par Runaway stars are another type of high-velocity stars, their velocity often point away from an OB association. One mechanism for producing a runaway star is to be ejected in binaries when its companion has exploded as a supernova \citep{blaauw61origin,tauris98Runaway,portegieszwart00Characteristics,wang09Companion,wang13Producing}. \citet{tauris15Maximum} investigated the ejection velocity of early and late type stars under this mechanism and they found that the upper limit of the ejection velocity of B-stars ($\sim 10 M_\odot$) can reach 550 km s$^{-1}$ in the Galactic rest frame and that late-type stars are even faster. The other mechanism is ejected from a stellar system by the dynamical interactions \citep{poveda67Runaway}. \citet{gvaramadze09origin} explored the exchange encounters in the three-body system and found that the ejected star can obtain velocities of $\sim 200 - 400$ km s$^{-1}$, depending on their mass. The discoveries of HVS US 708 \citep{geier15fastest} and HD 271791 \citep{heber08Btype} provide strong support of the two mechanisms.  

\par Besides the mechanisms mentioned above, high-velocity stars could also be ejected by other mechanisms such as from the tidal debris of disrupted dwarf galaxies or gravitational slingshot effect \citep[e.g.,][]{boubert16Dipole,garcia-bellido17Massive,montanari19Searching}. \citet{abadi09Alternative} used simulations to predict HVS can be ejected from tidal disrupted dwarf galaxies. A close interaction between globular cluster and a single supermassive black hole or a binary supermassive black hole could lead to the ejection of high-velocity stars \citep{capuzzo-dolcetta15Highvelocity,fragione16Highvelocity}. The HVS HE 0437-5439 is suggested to be ejected from Large Magellanic Cloud (LMC) through Hills mechanism \citep{erkal19hypervelocity}. \citet{du18Origin,du19New} use Gaia and LAMOST data to discover that some high-velocity stars origin from the tidal debris of disrupted dwarf galaxies. \citet{caffau20Highspeed} used the Gaia DR2 and VLT data to present the chemodynamical investigation of a 72 high-velocity sample stars and suggested that these stars can be the result of a disrupted small galaxy or  globular cluster members. \citet{marchetti21Gaia} derived that 7 high-velocity stars possibly origin from an extragalactic origin from Gaia EDR 3. All these results provide strong evidences that high-velocity stars origin from accreted and disrupted dwarf spheroidal galaxies. But most studies could not judge which dwarf galaxy or globular cluster these high-velocity stars origin from. Recently, \citet{huang21Discovery} discover a candidate HVS (J1443+1453) with a total velocity of 559 km s$^{-1}$ that is probably from Sgr.  

\par In this letter, we report the discovery of 60 candidate high-velocity stars, probably originated from the Sgr. Among them, 2 HVS candidates with extreme velocity that could be produced by the Hills mechanism, requiring a central MBH in the Sgr. This letter is organized as follows: in Section \ref{sec:samples}, we describe the method of assembling sample high-velocity stars. Section \ref{sec:orbit} explore the possible origin of sample stars by orbital analysis. Finally, we give a summary and discussion in Section \ref{sec:summary}.

\section{samples}
\label{sec:samples}

\par Gaia Early Data Release 3 \citep[Gaia EDR3,][]{collaboration21Gaia} provides positions, proper motions, and parallaxes for about 1.468 billion sources. It also contains radial velocities for about 7.21 million stars from Gaia DR2 \citep{collaboration18Gaia}. Furthermore, we cross-match Gaia EDR3 catalog with the Gaia-ESO DR4 \citep{gilmore12GaiaESO,randich13GaiaESO}, the Galactic Archaeology with HERMES (GALAH) DR3 \citep{desilva15GALAH,buder21GALAH}, low and medium resolution spectra (LRS and MRS) of Large Sky Area Multi-object Fiber Spectroscopic Telescope (LAMOST) DR9\citep{zhao12LAMOST,cui12Large,liu20LAMOST}, the Radial Velocity Experiment (RAVE) DR6 \citep{steinmetz20Sixth,steinmetz20Sixtha}, the Sloan Digital Sky Surveys (SDSS) optical spectra DR17 \citep{york00Sloan,yanny09SEGUE,abdurrouf22Seventeenth}, the Apache Point Observatory Galactic Evolution Experiment \citep[APOGEE,][]{majewski17Apache} of SDSS DR17 and the Southern Stellar Stream Spectroscopic Survey ($S^5$) DR1 \citep{li19southern} to supplement radial velocity information. 

\par We select stars with \texttt{RUWE}$<1.4$, where \texttt{RUWE} (Renormalised Unit Weight Error) is a reliable and informative goodness-of-fit statistic in Gaia EDR3 \citep{lindegren21Gaia}. We use \citet{lindegren20Gaia} method to correct the Gaia EDR3 parallaxes ($\varpi$) and remove stars $\varpi < 5\sigma_\varpi$. For spectroscopic data, we require S/N$>10$.

\par We use the median value of radial velocity ($rv$) difference $\delta v = rv_\mathrm{survey} - rv_{Gaia}$ to correct the $rv$ offset of other surveys relative to Gaia and the results are listed in Table~\ref{tab:correction}.

\par The misestimation of radial velocity error is reported by many study \citep[e.g.,][]{cottaar14INSYNC,tsantaki22Survey}. Following the method outlined in \citet{li19southern}, we use the mixture model
\begin{equation}
\label{equ:mixture}
\delta v_{i,j} \sim f \mathcal{N} \left(0, \sqrt{F^2(\sigma_{v,i}) + F^2(\sigma_{v,j})} \right) + (1-f) \mathcal{N} (0, \sigma_\mathrm{outl}),
\end{equation}
where $\delta v_{i,j}$ is pairwise radial velocity differences, $f$ is the weight of first mixture component, $\sigma_{v,i}$ and $\sigma_{v,j}$ are the radial velocity error of the $i$-th and $j$-th observation respectively and $\sigma_\mathrm{outl}$ is the outlier of radial velocity difference. The radial velocity correction function is $F(\sigma_{v}) = \sqrt{(k \times \sigma_v)^2 + \sigma_{v, \mathrm{floor}}^2}$, where $k$ is correction factor and $\sigma_{v, \mathrm{floor}}$ is the floor error of radial velocity.

\begin{deluxetable}{lCCCc}
\label{tab:correction}
\tablecaption{Correction parameters for radial velocity and its error.}
\tablehead{
	\colhead{Survey} & \colhead{offset} & \colhead{$k$} & \colhead{$\sigma_{v, \mathrm{floor}}$} & \colhead{Method}\\
	\colhead{} & \colhead{km s$^{-1}$} & \colhead{} & \colhead{km s$^{-1}$} & \colhead{}
}
\decimals
\startdata
LAMOST LRS &  -5.11 &   0.81 &         0.0 &   dupl \\
LAMOST MRS &  -0.82 &   0.99 &        1.78 &   dupl \\
APOGEE     &  -0.09 &   1.78 &        0.08 &   dupl \\
RAVE       &  -0.32 &   0.92 &        0.89 &   dupl \\
SDSS       &   7.69 &   1.41 &         2.2 &   dupl \\
S$^5$      &  -0.35 &      - &           - &      - \\
GALAH      &  -0.32 &   2.92 &        0.31 &  cross \\
Gaia-ESO   &   -0.3 &   3.33 &         0.0 &  cross \\
Gaia       &      - &   1.56 &         0.3 &  cross \\
\enddata
\tablecomments{Column 1 is the survey name; Column 2 lists the radial velocity offset from Gaia; Columns 3-4 are the correction parameters applied in Equation~\ref{equ:mixture}; Column 5 gives the correction method used.}
\end{deluxetable}

\par For surveys with at least 10$^4$ sources that contain multiple observations, we fit this model to multiple observations. The results are listed in Table~\ref{tab:correction} and the correction method of these surveys in Table ~\ref{tab:correction} is set to ``dupl''. In addition, the radial velocity error in S$^5$ catalog has been corrected based on this method \citep{li19southern}.

\par We apply the similar method to other surveys. In Equation~\ref{equ:mixture}, the $i$-th observation is from only one of the uncorrected surveys, and the $j$-th observation is from all corrected surveys. The results are also listed in Table~\ref{tab:correction} and the method is marked as ``cross''.

\par After applying the above corrections, we weight the radial velocity means by the inverse of its variance. Observations that deviate from the average by more than $3\sigma$ are removed. We only retain stars with the epoch number of radial velocity $n_{rv} \geqslant 4$ to ensure that the radial velocity can be considered as the systemic velocity of the system \citep{boubert19Lessons}.

\par To search for stars that originated from dwarf galaxies or globular clusters in our sample, we collect 50 dwarf galaxies and 160 globular clusters. 46 dwarf galaxies are referenced from \citet{li21Gaia}. We add Leo T, Pegasus III, Phoenix I \citep{mcconnachie20Updated} and Sgr \citep{fritz18Gaia} to our dwarf galaxy sample. All 160 globular clusters are referenced from \citet{vasiliev21Gaia}.  

\par Heliocentric distances ($d$) and velocities in right ascension and declination direction ($v_{\alpha}$ and $v_{\delta}$) are derived using the Bayesian method \citep{luri18Gaia}. We use the posterior of \citet{du19New} and replace the distance prior with the three-parameter Generalized Gamma Distribution \citep{bailer-jones21Estimating}. The priors for both $v_{\alpha}$ and $v_{\delta}$ are uniform. We use the Markov chain Monte Carlo (MCMC) sampler, \texttt{emcee} \citep{foreman-mackey13emcee}, to draw samples from the posterior probability.  The distances and velocities of dwarf galaxies and globular clusters are calculated using the similar method. Due to high accuracy of their distance modulus and distance, we use uniform distributions as distance priors. In this study, we generate 2000 Monte Carlo (MC) realizations for each object (stars, dwarf galaxies and globular clusters) and use the median value to describe the results. The lower and upper uncertainties are the 16th and 84th percentiles of the probability distribution function.

\par We use the right-handed Cartesian Galactocentric coordinate system \citep{juric08Milky}. Here we adopt the distance of the Sun to the GC $R_{\odot} = 8.122$ kpc \citep{gravitycollaboration18Detection} and the vertical distance from the Sun to the Galactic midplane $z_{\odot} = 25$ pc \citep{juric08Milky}. The proper motions, radial velocities and distances are used to derived 3D velocities ($U$, $V$, $W$) in the right-handed coordinate system \citep{johnson87Calculating}.

\par To get a accurate high-velocity star sample, we select stars with $\sigma_{v_\mathrm{GC}}/ v_\mathrm{GC}  <  30\% $ and $v_\mathrm{GC} > 0.75 v_\mathrm{esc} (\vec{r})$, where $v_\mathrm{GC}$ and $\sigma{v_\mathrm{GC}}$ are the Galactic total velocity and its error, $v_\mathrm{esc} (\vec{r})$ is the escape velocity at $\vec{r} = (x,y,z)$ in \citet{eilers19Circular} potential model, which includes a Plummer \citep{binney87Galactic} bulge, two Miyamoto-Nagai \citep{miyamoto75Threedimensional} disks and a NFW \citep{navarro97Universal} dark matter halo. By applying the above selection criteria, we obtain 1049 sample stars.

\section{orbit analysis}
\label{sec:orbit}

\par To get some hints on the ejection location of our sample high-velocity stars, we study their orbit by adopting the potential model in \citet{eilers19Circular}
We use \texttt{galpy} \citep{bovy15galpy} to get the orbit of our Bayesian samples (stars, dwarf galaxies and globular clusters), and integrate orbit back in a total time of 1 Gyr. The gravity influence of dwarf galaxy and globular cluster is temporarily ignored.

\par We define the normalized distance from the star to the dwarf galaxy and to the globular cluster:
\begin{equation}
\Delta r_\mathrm{affi}^* = 
\begin{cases}
	\frac{|\vec{r}_\mathrm{star} - \vec{r}_\mathrm{dwarf}|}{r_\mathrm{half}}, &\mathrm{for\ dwarf\ galaxies} \\
	\frac{|\vec{r}_\mathrm{star} - \vec{r}_\mathrm{cluster}|}{R_0}, &\mathrm{for\ globular\ clusters,}
\end{cases}
\end{equation}
where $\vec{r} = (x,y,z)$ is Galactocentric coordinates, $d$ is the Heliocentric distance, $r_\mathrm{half}$ is the half-light radius and $R_0$ is the scale radius of the Plummer model \citep{vasiliev21Gaia}. Most of $r_\mathrm{half}$ are referenced from \citet{mcconnachie20Revised} except Crater I \citep{laevens14New}. The relative minimum value of $\Delta r_\mathrm{affi}^*$ is marked as $\Delta r_\mathrm{min}^*$. If there are multiple relative minima, we take the nth which maximizes the probability that the star originated from the dwarf galaxy or the globular cluster $P_\mathrm{affi}$. This probability is defined as the fraction of MC realizations with $\Delta r_\mathrm{min}^* <2.5$. 

\par There are 60 stars with $P_\mathrm{Sgr}>50\%$, which means the probability of these stars originating from Sgr is greater than 50\%. We have not found any stars with probability of originating from other dwarf galaxies or globular clusters greater than 50\%. The maximum probability of originating from non-Sgr objects is 11.4\% and this possible object is Antlia II with a half-light radius of 2.95 kpc. The maximum probability of that a star originates from globular clusters is 6.3\%, and the possible source is $\omega$ Cen with a scale radius of 23 pc. To test the influence of Sgr's large half-light radius, we create a ``mirror'' Sgr by reversing the sign of Sgr's $z$ and $v_z$. Base on the same method as above, we find 24 stars with $P_\mathrm{mirror}>50\%$, which is less than half of the number of stars originating from Sgr.

\par In order to further determine the link between these stars and Sgr, we use \texttt{GIZMO} \citep{springel05cosmological,hopkins15new} to take the gravity influence of Sgr into account by assuming a Keplerian potential with a total mass of $4 \times 10^8 M_\odot$ for Sgr \citep{vasiliev20last}. We select 215 stars with $\Delta r_\mathrm{min}^* <5.5$ relative to Sgr, and repeat the above orbit analysis in a total time of 0.5 Gyr. The change of $\Delta r_\mathrm{min}^*$ is very small, equivalent to several parsecs. This is approximately equal to the half-light radius of most globular clusters, therefore globular cluster $\Delta r_\mathrm{min}^*$ is very sensitive to the gravitational potential. The probability of stars originating from a globular cluster is therefore greatly affected. But dwarf galaxies ($r_\mathrm{half} \sim 100$ pc) are hardly affected.

\par The Large Magellanic Cloud (LMC) with a mass of $1.38 \times 10^{11} M_\odot$ \citep{erkal19total} is an important gravitational source in the Milky Way. Its proper motions are referenced from \citet{gaiacollaboration20Gaia} and other parameters are from \citet{mcconnachie20Updated}. We also use the Bayesian method to obtains 2000 random samples. Because of its distance, it is considered a point source. For most of our selection of 215 stars, the change of $\Delta r_\mathrm{min}^*$ is less than 100 pc compared to only considering the Sgr's gravity. $\Delta r_\mathrm{min}^*$ of dwarf galaxies with smaller half-light radius may be significantly affected. The change of $\Delta r_\mathrm{min}^*$ for 8 stars is more than 1 kpc, and the maximum value is 7.9 kpc.

\par
\begin{figure}
\plotone{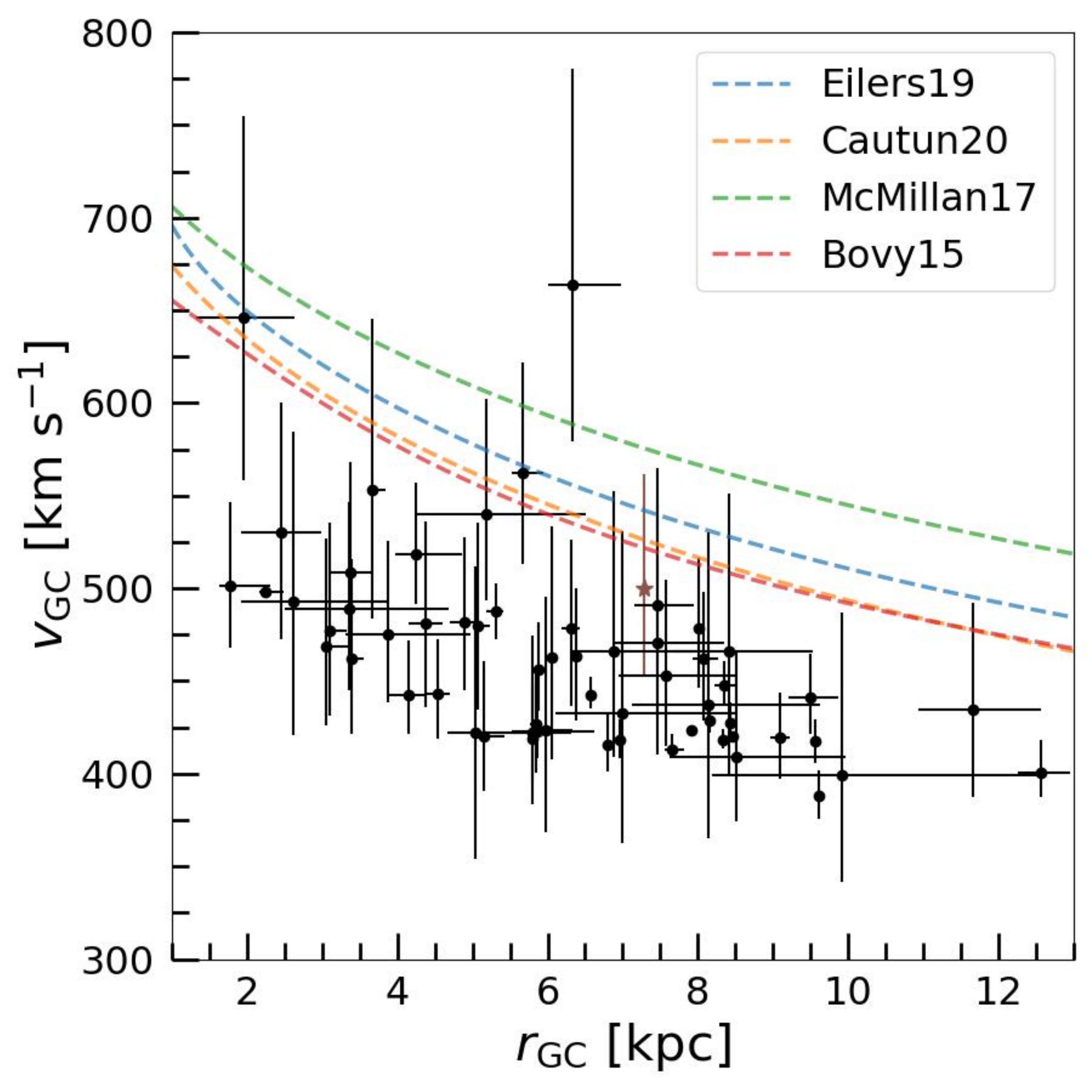}
\caption{Distribution of total velocity $v_\mathrm{GC}$ versus Galactocentric distance $r_\mathrm{GC}$ for stars with $P_\mathrm{Sgr} >50\%$. Black points with errorbars represent stars with the probability of correlation with Sgr $P_\mathrm{Sgr} > 0.5$. Brown star with errorbars is the candidate star that probably from Sgr found by \citet{huang21Discovery}. Four different colored dashed lines are escape velocity curves derived from different Milky Way potential models \citep{eilers19Circular,cautun20milky,mcmillan17mass,bovy15galpy}.}
\label{fig:escape}
\end{figure}

 Finally, there are 60 stars with $P_\mathrm{Sgr}>50\%$. The distribution of these stars in $v_\mathrm{GC}$ and $r_\mathrm{GC}$ space is shown in Figure \ref{fig:escape}. We also plot the escape velocity curve of four Milky Way potential models \citep{eilers19Circular,cautun20milky,mcmillan17mass,bovy15galpy} as a function of Galactocentric distance in this figure.

\begin{figure*}
\gridline{\fig{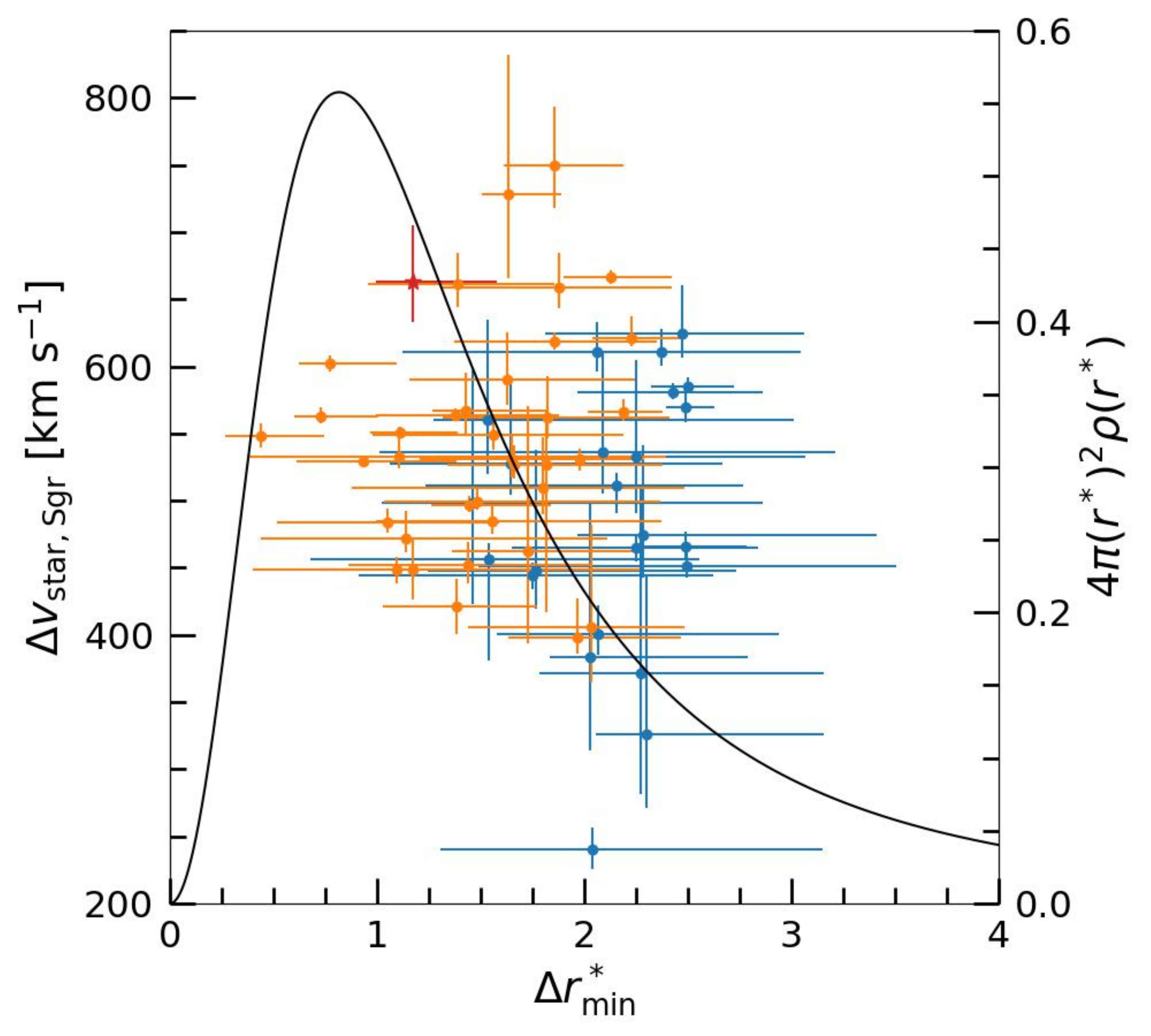}{0.475\textwidth}{(a)}
		\fig{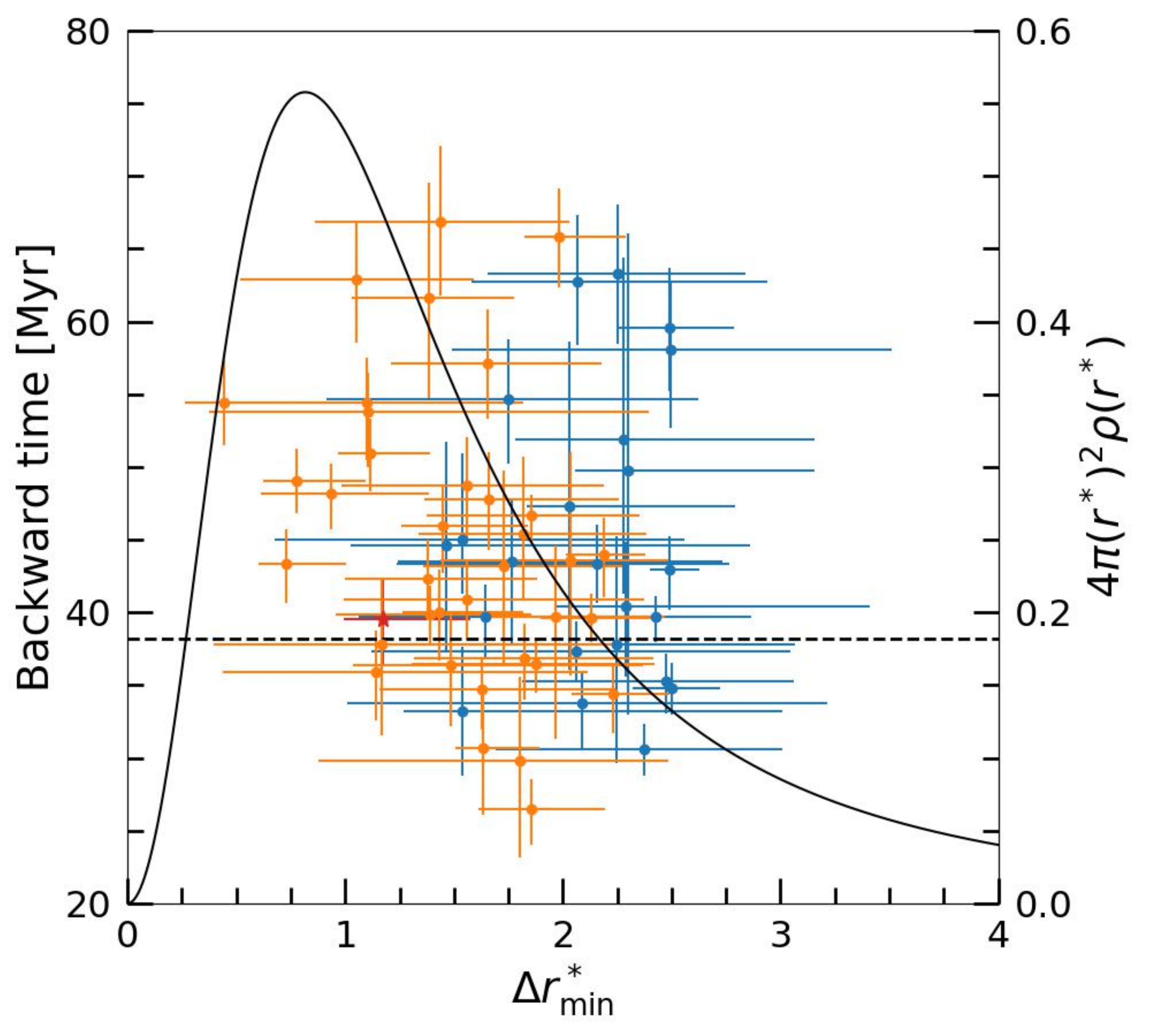}{0.475\textwidth}{(b)}
}
\caption{Panel (a): Distribution of the velocity and the half-light radius normalized distance of stars relative to Sgr when they are at the closest point. Stars with the probability $P_\mathrm{Sgr} >50\%$ originating from the Sgr are shown in blue points. Orange points represent stars with a probability of $P_\mathrm{Sgr} >84\%$. The Black line shows the density at the spherical shell as a function of the half-light radius normalized distance from the Sgr center. The candidate star that probably from Sgr found by \citet{huang21Discovery} is shown in red star. Panel (b): The backward time versus the normalized distance between the star and the Sgr of the closest approach. The symbols are same as in the panel (a). The black horizontal dashed line shows the backward time when the Sgr arrives its pericenter.}
\label{fig:dr_dwarf}
\end{figure*}

\par Figure \ref{fig:dr_dwarf} shows the properties of stars at the time of the closest approach. In the panel (a), we show the distribution of velocity difference versus half-light radius normalized distance between the star and the Sgr when they are the closest approach. We consider the velocity of the stars relative to Sgr as their ejection velocity $v_\mathrm{ej}$. There are 2 stars with $v_\mathrm{ej} \sim 740$ km s$^{-1}$, such a fast velocity may be produced by the Hills mechanism \citep{hills88Hypervelocity,bromley06Hypervelocity}. The probability of an ejection is
\begin{equation}
P_\mathrm{ej} \approx
\begin{cases}
1- D_\mathrm{min}/175, &\mathrm{if\ 0 \leqslant D_\mathrm{min} \leqslant 175} \\
0, &\mathrm{otherwise,}
\end{cases}
\end{equation}
where
\begin{equation}
D_\mathrm{min} = \frac{R_\mathrm{min}}{a_\mathrm{bin}} \left[\frac{2M_\mathrm{bh}}{10^6 (m_1+m_2)}\right]^{-1/3},
\end{equation}
where $R_\mathrm{min}$ is the distance between the binary and the black hole when the close encounter, $a_\mathrm{bin}$ is the semi-major axis of the binary, $M_\mathrm{bh}, m_1, m_2$ are masses of the black hole and the binary, respectively. However, $R_\mathrm{min}$ (i.e., $\Delta r_\mathrm{min}$ in our study) is much larger than a reasonable binary semi-major axis for our candidate stars. This indicates $D_\mathrm{min} > 175$, so $P_\mathrm{ej} = 0$. This does not mean that the Hills mechanism is not a reasonable explanation for the candidate Sgr HVSs, because the orbital uncertainty could increase $D_\mathrm{min}$ to more than 175. The small error in $\Delta r_\mathrm{min}$ suggests that statistical error is not the main reason. Additional gravitational influences may cause changes in the orbits of stars after they are ejected, such as the globular clusters associated with Sgr \citep[][NGC 6715 (M54), Terzan 7, Arp 2 and Terzan 8]{ibata95Sagittarius} and Sgr stream. Therefore, the mass of the black hole cannot be estimated, which needs to be further studied by N-body simulations in the future.

\par Another ejection mechanism of high-velocity stars is the tidal disruption of the dwarf galaxy \citep{abadi09Alternative}. Their simulation shows that during the last time the dwarf galaxy passed through the pericenter, the velocity distribution of the stars in the tidal debris extends beyond the escape velocity. As shown in the panel (b) of Figure \ref{fig:dr_dwarf}, the backward time $t_\mathrm{trace}$ of the closest approach and the Sgr passed its pericenter are close for 59 stars with $P_\mathrm{Sgr} >50\%$ (except one star with $t_\mathrm{trace} \sim 150$ Myr). This is consistent with the results from \citet{abadi09Alternative}. With the release of more high-precision data, we could discover more high-velocity stars from the accreted and disrupted dwarf spheroidal galaxies, which would help us better understand the property of dwarf galaxies and the history of the Milky Way’s accretion.

\begin{figure}
\plotone{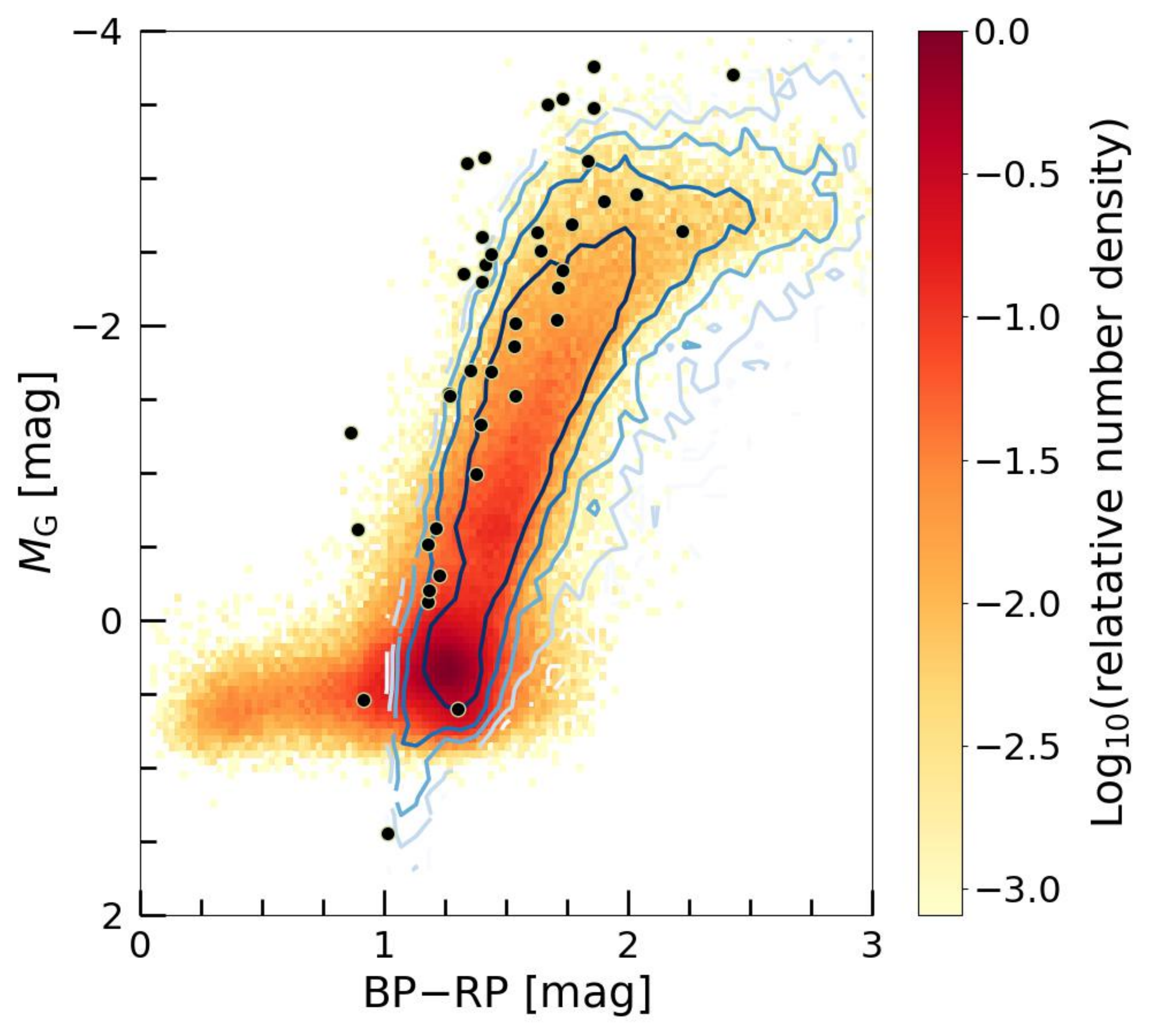}
\caption{Color–magnitude diagram of stars with probability originating from Sgr greater than 50\% (black points) compared to the Sgr \citep[red pixels,][]{vasiliev20last} and the Sgr stream \citep[blue contour lines,][]{vasiliev21Tango}. The color of the pixel is coded by the logarithmic number density relative to the maximum density. The contour lines are equal logarithmic number density lines with an interval of 0.5 dex.}
\label{fig:mag}
\end{figure}

\par Figure~\ref{fig:mag} shows the Hertzsprung-Russell (HR) diagram for stars with the probability $P_\mathrm{Sgr} >50\%$. The $x$-axis is the color index in the Gaia BP-band and RP-band. The $y$-axis is the Gaia G-band absolute magnitude, which is corrected by the extinction provided by \citep{green193D}. All of them are late-type giant stars. As shown in this figure, these Sgr high-velocity stars are most similar to both Sgr and Sgr stream, but they are closer to Sgr stream relative to Sgr which is more concentrated at ($M_\mathrm{G}$, BP$-$RP) = (0.5, 1.25). This suggests that these stars have a similar origin to Sgr stream.

\begin{figure}
\plotone{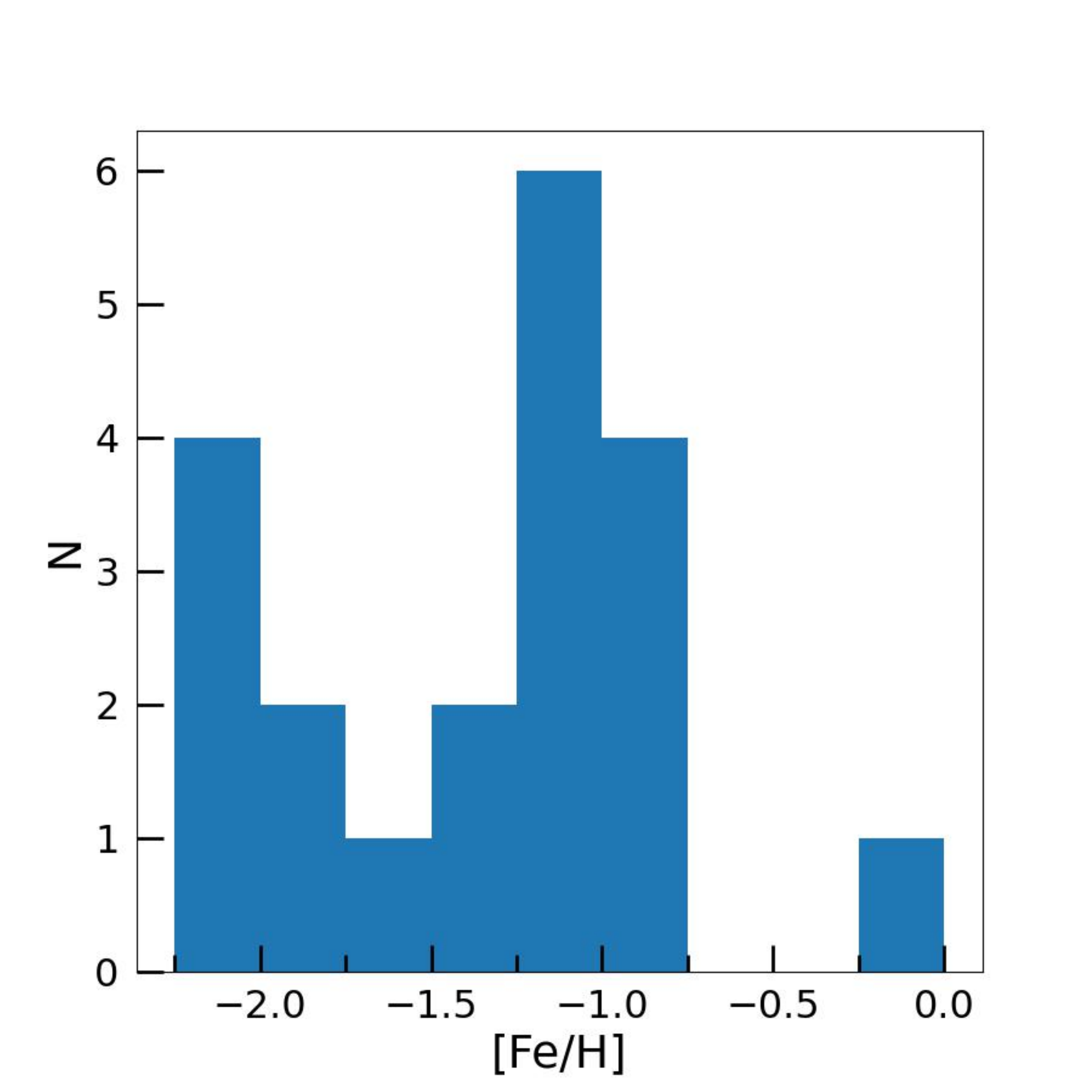}
\caption{Histogram of metallicity distribution of stars with probability originating from Sgr greater than 50\%.}
\label{fig:feh}
\end{figure}  

\par Of the 60 stars with $P_\mathrm{Sgr} >50\%$, only 19 have metallicity information, the distribution is shown in the Figure~\ref{fig:feh}. There are two peaks in this distribution. The first one is at [Fe/H]$\sim -1$, which is very close to the metallicity distribution of the Sgr stream \citep{hayes20Metallicity}. This is consistent with the assumption that they are tidally stripped stars from the Sgr. The peak at [Fe/H]$\sim -2$ is close to the most metal-poor part of the Sgr stream, and they may originate from the Sgr halo. One metal-rich star ([Fe/H]$=-0.03$) reaches the upper limit of the Sgr's metallicity, and has a retrograde orbit. It is located at ($M_\mathrm{G}$, BP$-$RP) = (0.6, 1.3) in Figure~\ref{fig:mag}, very close to the densest region of Sgr ($M_\mathrm{G}$, BP$-$RP) = (0.5, 1.25) and may be related to the most metal-rich part of Sgr.

\par Recently, \citet{huang21Discovery} discovered a candidate high-velocity star originating from Sgr. It is not included in our sample because its radial velocity based only on two epochs. We test this star using our method and find similar results. It is shown in Figure \ref{fig:escape} and Figure \ref{fig:dr_dwarf}. Our determined minimum distance $\Delta r_\mathrm{min} = 3.33_{-0.51}^{+1.09}$ kpc, the trace back time $t_\mathrm{trace} = 39.4_{-3.2}^{+2.6}$ Myr and the velocity difference $\Delta v = 665_{-30}^{+43}$ km s$^{-1}$ correspond to their results: $\Delta r_\mathrm{min} = 2.42_{-0.77}^{+1.8}$ kpc, $t_\mathrm{trace} = 37.8_{-6.0}^{+4.6}$ Myr and $\Delta v = 690_{-65}^{+104}$ km s$^{-1}$. This difference has little effect on dwarf galaxies with a large half-light radius like Sgr, but it may have effect for some dwarf galaxies with a small half-light radius.

\section{summary and discussion}
\label{sec:summary}

\par In this letter, based on the high-quality positions, proper motions and parallaxes of Gaia EDR3 combining with the radial velocity of several surveys, we define a high-velocity star sample with $v_\mathrm{GC}> 0.75 v_\mathrm{esc}$. By analyzing backward orbits, we investigate the links between these stars and 50 dwarf galaxies, 160 globular clusters. Finally, there are 60 stars that had a closest approach with Sgr and the encountering time is close to when Sgr reached its pericenter and all of them are late-type giants. A catalog of the properties of these stars is available at \url{https://nadc.china-vo.org/res/r101121/.}

\par The effect of the Sgr's gravity can be ignored. Even further considering the gravity of the LMC, very few stars are significantly affected. The similarity of the 60 stars and Sgr stream in the HR diagram suggests they could from the Sgr. 19 of these stars with available [Fe/H] are also chemically similar to Sgr stream. Moreover, we found 2 hypervelocity stars with an ejection velocity $v_\mathrm{ej} \sim 740$ km s$^{-1}$. This extreme velocity can be produced by the Hills mechanism \citep{hills88Hypervelocity}, but it is difficult to constrain the properties of black holes at the center of the Sgr.

\par Because Sgr has just passed its pericenter ($\sim$ 38.2 Myr ago), we could expect to discover more high-velocity stars originated from Sgr with the release of more and more large surveys data. The details, such as number and velocity distribution, need to be studied using numerical simulations. This can help us study the interaction between Sgr and the Milky Way in the accretion event, which will provide important clues to the formation and evolution history of the Milky Way galaxy. On the other hand, most of the dwarf galaxies discovered are located near their pericenter \citep{li21Gaia}. This means that we can better study the high-velocity stars of tidal origin, and it is helpful to understand different ejection mechanisms of high-velocity stars.

\begin{acknowledgments}

\par We thank especially the referee for insightful comments and suggestions, which have improved the paper significantly. This work was supported by the National Natural Foundation of China (NSFC No. 11973042, No. 11973052 and No. 11873053).  It was also supported by the Fundamental Research Funds for the Central Universities and the National Key R\&D Program of China No. 2019YFA0405501. H.J.N. acknowledges funding from US NSF grant AST-1908653. This project was developed in part at the 2016 NYC Gaia Sprint, hosted by the Center for Computational Astrophysics at the Simons Foundation in New York City.

\par This work has made use of data from the European Space Agency (ESA) mission {\it Gaia} (\url{https://www.cosmos.esa.int/gaia}), processed by the {\it Gaia} Data Processing and Analysis Consortium (DPAC, \url{https://www.cosmos.esa.int/web/gaia/dpac/consortium}). Funding for the DPAC has been provided by national institutions, in particular the institutions participating in the {\it Gaia} Multilateral Agreement.

\par This work made use of the Third Data Release of the GALAH Survey. This paper includes data that has been provided by AAO Data Central (datacentral.aao.gov.au). It also made use of the Sixth Data Release of the RAVE Survey. Guoshoujing Telescope (the Large Sky Area Multi-Object Fiber Spectroscopic Telescope LAMOST) is a National Major Scientific Project built by the Chinese Academy of Sciences. Funding for the project has been provided by the National Development and Reform Commission. Funding for the Sloan Digital Sky Survey IV has been provided by the Alfred P. Sloan Foundation, the U.S. Department of Energy Office of  Science, and the Participating  Institutions. SDSS-IV acknowledges support and resources from the Center for High Performance Computing at the  University of Utah. The SDSS website is www.sdss.org. Based on data acquired at the Anglo-Australian Telescope. We acknowledge the traditional owners of the land on which the AAT stands, the Gamilaraay people, and pay our respects to elders past and present.

\end{acknowledgments}

\bibliography{Library}
\bibliographystyle{aasjournal}


\end{document}